\documentclass{article}

\usepackage{arxiv}

\usepackage[T1]{fontenc}     
\usepackage{hyperref}        
\usepackage{url}             
\usepackage{booktabs}        
\usepackage{amsfonts}        
\usepackage{nicefrac}        
\usepackage{microtype}       
\usepackage{lipsum}          

\usepackage{graphicx}
\usepackage{amsthm}
\usepackage{amsmath}
\usepackage{cite}

\title{Advanced analytical method based on Green's theorem for light transmission through subwavelength structures of multiple configurations in metal films}

\date{} 					

\author{
  Jian-Shiung Hong \\
  Department of Physics\\
  National Cheng Kung University\\
  1 University Road, Tainan 70101, Taiwan, Republic of China \\
  \texttt{hongjs@phys.ncku.edu.tw} \\
   \And
  Kuan-Ren Chen \\
  Department of Physics\\
  National Cheng Kung University\\
  1 University Road, Tainan 70101, Taiwan, Republic of China \\
  \texttt{chenkr@phys.ncku.edu.tw} \\
}

\begin{document}
\maketitle

\begin{abstract}
Nowadays, methods for analyzing light transmission through subwavelength structures are typically based on the mode expansion with Fourier series. However, these methods require sophisticated techniques and the solutions are in $k$-space, where the coupling physics that associates the boundary field in real space and the structure geometry becomes obscure. Moreover, the typical methods for the analysis of multi-layered hybrid configurations can be exhaustive due to the complex mode couplings at the interfaces of the layers. In contrast, an early method can analyze the single-slit transmission for solutions entirely in real space [F. L. Neerhoff and G. Mur, Appl. Sci. Res. 28, 73 (1973)] by rigorously formulating the field based on Green's theorem and obtaining two types of the Green's function for the cylindrical wave mode in free space and the symmetric waveguide modes inside the slit, respectively. In this article, we advance the method by developing a new type of Green's function for the asymmetric waveguide modes inside a groove. Since these wave modes are independent in real space, the coupling physics from the method becomes straightforward and the solutions are intuitive. With this meticulous method, we show that complex and multi-layered hybrid configurations can be easily analyzed with excellent accuracy. In addition, the method demonstrates the capability of translating the wave interaction problems to analytical physical models for further interpretation and study.
\end{abstract}


\section{Introduction}
\label{s1}
The interaction between the electromagnetic (EM) wave and the subwavelength structures in metal films has been an active research area for decades due to the remarkable phenomena in the wave mechanics demonstrated, such as the extraordinary transmission through a 2D array of small holes \cite{Ebbesen, Haitao, Frerik} and the absorption anomalies in the reflection and transmission spectra of 1D metallic gratings of a small period \cite{Garcia1}. In addition, a predefined structure is capable of guiding or localizing the EM wave \cite{Maier}, or scattering it into space to be coupled to another structure \cite{Hutter}. Therefore, besides the academic importance in optical physics, the study opened up many potential applications, such as plasmonic circuit \cite{Gramotnev}, biosensing \cite{Anker}, plasmonic solar cell \cite{Atwater}, optical antennas \cite{Novotny}, optical data storage \cite{Challener}, and plasmonic lithography \cite{Fang}.

The transmission through various configurations of 1D subwavelength structures was extensively studied to understand the underlying physics of the wave mechanics and to manipulate the EM wave. For instance, the incident energy can be funneled and transmitted through a metallic slit of subwavelength scale \cite{Takakura, Sturman, Chang3, Li4, Hong1} to be concentrated at the exit, which implies the possibility of becoming the light source of a microscopy with superresolution in the near region \cite{Betzig, Kukhlevsky2, Kukhlevsky3, Kelso} or enhancing the signal intensity of a Raman spectroscopy \cite{Mechler}. In advance, a slit patterned with grooves can directionally beam or focus the light by controlling the geometric parameters or the surrounding medium \cite{Lezec, Martin, Garcia3, Lopez2, Shi1, Shi2, Chen3, Baron, Kim2, Hong2}; a double slit can be used to generate the coherence light source \cite{Weiner} and an indented double slit can focus the light beyond the diffraction limit in the intermediate zone \cite{Chen1, Chen2} or enhance the transmission \cite{Goncharenko}.

Therefore, methods for the analytical study of the wave interaction problem with the subwavelength structures and the design of the applicable devices are essential. Typically, the methods of expanding the particular modes to the specific boundary with Forious series \cite{Kreyszig} to solve the wave equations \cite{Petit} for the configurations of a single slit \cite{Takakura, Bravo}, a slit patterned with grooves \cite{Martin, Garcia2, Villate2, Na2}, and a double slit \cite{Gordon2} were commonly seen. In these methods, all modes expanded are internally coupled in $k$-space, while the coupling of the field between the openings was often described as mode projection \cite{Bravo, Martin, Garcia2, Villate2}. To obtain the solutions to the EM field, the methods usually required the mode matching technique based on the Fourier transform of the eigenfunctions \cite{Takakura, Na2} or the effecttive surface impedance boundary conditions \cite{Bravo, Martin, Garcia2, Villate2, Gordon2}. However, since the solutions are in $k$-space, the coupling physics that associates the boundary field in real space and the structure geometry becomes obscure. In the meantime, the mathematical treatment of the method may cause a great barrier for unexperienced researchers in studying the wave interaction problem analytically. Moreover, the typical methods for multi-layered hybrid configurations, e.g., an indented double slit \cite{Chen1, Chen2, Goncharenko}, can be exhaustive due to the complex couplings at the interfaces of the layers \cite{Gottmann, Young}.

The Green's theorem has been successfully employed to study the conventional diffraction problem, in which the wave field is considered scalar \cite{Goodman}, with remarkable accuracy. Few decades ago, Neerhoff and Mur \cite{Neerhoff} revisited the scalar theory to study the diffraction problem by a wide metallic slit of finite thickness. In this method, the solution was entirely in real space. First, the theorem yielded the rigorous formulation of the field in the system. Second, two types of the Green's function were obtained from the special solutions \cite{Kreyszig} (in this case, the Hankel function of the first kind) to standard point source radiation \cite{Goodman} in a bounded space and the method of images \cite{Morse} to describe the cylindrical wave mode in free space and the symmetric waveguide modes inside the slit, respectively. Betzig et al. \cite{Betzig} confirmed the feasibility of the method for a single slit of width much smaller than the wavelength. Kukhlevsky et al. compared the method with the Rayleigh's expansion \cite{Kukhlevsky4}, characterized the transmission and reflection \cite{Kukhlevsky}, studied the transmission using a fs-scale pulse \cite{Kukhlevsky2, Kukhlevsky3, Mechler2}, or analyzed a tip-shaped slit transmission to enhance the signal intensity \cite{Mechler}. Another approach that applies the theorem was given with the formulated linear equations mixed with the boundary integral and the boundary conditions \cite{Kelso, Lindberg}. However, the application of the Green's theorem as a more general analytical method was underestimated such that it has never been employed for configurations other than a single slit so far.

In this article, we advance the Neerhoff and Mur's method and show that, besides the single slit, it is also capable of analyzing the light transmission through subwavelength structures of more complex configurations, including a slit patterned with grooves, a double slit, and an indented double slit. Without loss of generality, we assume that the structure is made of a perfect electric conductor (PEC). This method aims at solving the scalar phasor of the transverse magnetic field with the time harmonic dependence. With the Green's theorem, we formulate the field representations for each configuration and then reduce them to sets of simultaneous integral equations based on the boundary conditions for the solutions. In addition to the two types of the Green's function, we develop a new type for the wave modes inside a groove: while the two parallel boundaries of the groove confine the source radiation as a slit waveguide, this type describes asymmetric waveguide modes due to the additional reflection at the third boundary. These explicit Green's functions are then employed according to the boundary of the structures. Since these wave modes are independent in real space, the coupling physics from the method becomes straightforward: the sum of the fields that enter or leave the opening in the outside are equal to those that propagate away or toward the opening in the inside. The method therefore provides a simple framework and obtains solutions with intuitiveness.

For the first time, the complex configurations more than a single slit, including the slit patterned with grooves, the double slit, and the indented double slit, are analyze with this meticulous method; especially, we show that the indented double slit as a multi-layered hybrid configuration is easily included in the analysis. The results are all in excellent agreement with the finite-difference time-domain (FDTD) simulations. Moreover, since the method rigorously formulates the EM field, it demonstrates the capability of translating the wave interaction problem to physical models, e.g., Fabry-P\'{e}rot-like resonance for transmission through a single slit \cite{Chang3, Hong1, Li4} or point source interference for diffraction by a slit patterned with grooves \cite{Hong2}, for the further interpretation and study. In addition, it is discussed that this fundamental method may still be applicable when the plasmonic effect is considered.

We introduce the basic assumption for the EM field and the application of the Green's theorem in Sections \ref{s2} and \ref{s3}. The formulations and the solutions to the fields for a single slit, a slit patterned with grooves, a double slit, and an indented double slit are presented in Sections \ref{s4}-\ref{s7}. Section \ref{s8} discusses the translation of the wave interaction problem to physical models with the developed method. The summary is in Section \ref{s9}. \ref{sa} shows the solution to the fields for the single-slit configuration.

\section{Formulation of time-harmonic fields}
\label{s2}
\begin{figure}[htbp]
\centering\includegraphics{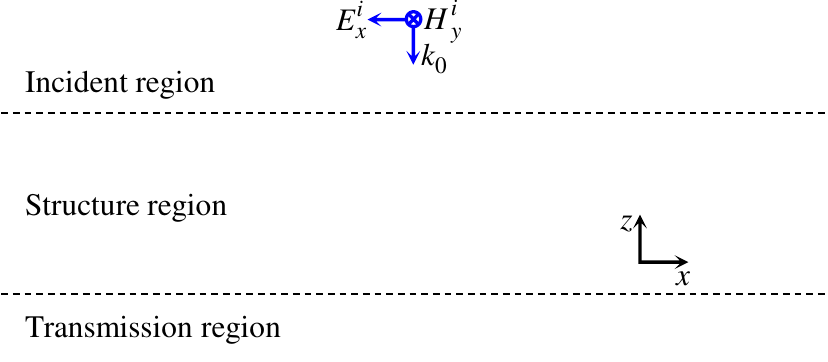}
\caption{Schematic of the incident region, the structure region, and the transmission region. A $p$-polarized wave propagates in the incident region toward the structure region.}
\label{Fig01}
\end{figure}

To analyze the transmission through subwavelength structures in a 2D system, we consider a $p$-polarized plane wave in the incident region that propagates in the $x$-$z$ plane toward the structure region to the transmission region, as depicted in Fig. \ref{Fig01}. The structure region can be represented by one or more sub-regions, depending on the configuration of interest; the coordinate origin then will be determined accordingly. In this article, we only consider the normal incidence; in addition, each region is filled with vacuum for simplicity. With the analytical method, we will obtain the EM field in all regions. However, it should be noted that the oblique incidence and dielectrics can be included in this method \cite{Betzig, Neerhoff}.

Without loss of generality, we also assume that the polarized magnetic field is time-harmonic and constant in the $ y $ direction:
\begin{equation} \label{eq01}
{\mathbf{H}}(x,y,z,t) = \hat y U(x,z) \exp(- i\omega t),
\end{equation}
where $ U(x, z) $ is the scalar phasor due to the polarization in the 2D system. Let the impedance of vacuum $\eta_0$ be normalized to unity. Once $U(x,z)$ is obtained, the electric field $\mathbf{E} = \hat x E_x + \hat y E_y + \hat z E_z$ is yielded as
\begin{equation} \label{eq02}
{E_x} = \frac{{ - i}}
{{k_0}}{\partial _z}U(x,z),{E_y} = 0,{E_z} = \frac{i}
{{k_0}}{\partial _x}U(x,z).
\end{equation}
We assume that the structure is made of a PEC. Since the tangential $ \mathbf{E} $ should vanish at the surface, we let 
\begin{equation} \label{eq03}
\partial_{n} U = 0 \quad {\text{at the conductor surface}},
\end{equation}
where $n$ is the normal vector at the surface.

In each main and sub-region numbered from the top to bottom, we represent the field as $U_{j}(x, z)$ $(j = 1, 2, 3, ...) $. The time harmonic behavior indicates that the field satisfies the Helmholtz equation:
\begin{equation} \label{eq04}
({\nabla ^2} + k_0^2){U_j} = 0.
\end{equation}
In the incident region, the field is further decomposed into three components:
\begin{equation} \label{eq05}
{U_1}(x,z) = {U^i}(x,z) + {U^r}(x,z) + {U^d}(x,z),
\end{equation}
each of which satisfies
\begin{equation} \label{eq06}
({\nabla ^2} + k_0^2){U^{i,r,d}} = 0.
\end{equation}
The scalar field $ U^{i} $ represents that of the incident wave, which, for simplicity, is assumed to be a plane wave of unit amplitude:
\begin{equation} \label{eq07}
{U^i}(x,z) = \exp (-ik_0 z).
\end{equation}
The scalar field $U^{r}$ denotes that of the reflected wave as if there were no opening at the interface between the incident and structure region, i.e., all the openings are short-circuited \cite{Butler, Harrington, Chang1, Haddadpour}. Therefore, it gives
\begin{equation} \label{eq08}
{U^r}(x,z) = \exp (ik_0 z + i\phi),
\end{equation}
where $\phi$ is the phase difference from the incident field. Suppose that the interface is at $z = z_{b}$. The boundary condition in Eq. (\ref{eq03}) for the tangential field $\mathbf{E}$ should apply. We obtain
\begin{equation} \label{eq09}
\partial_{z} U^{i}(x,z)|_{z = z_{b}} + \partial_{z} U^{r}(x,z)|_{z = z_{b}} = 0 \quad {\text{for }} -\infty < x < \infty.
\end{equation}
The above equation can only be valid for all $ x $ if $\phi = -2k_{0} z_{b}$. We then obtain
\begin{equation} \label{eq10}
{U^r}(x,z) = {U^i}(x,2{z_b} - z).
\end{equation}
The scalar field $U^{d}$ describes that of the diffracted wave in the incident region due to the presence of the opening(s).

Between two neighboring regions, the tangential $ \mathbf{H} $ and $ \mathbf{E} $ at the opening(s) of the structures are continuous. For the interface at $ z = z_{bj} $ between regions $ j $ and $ (j + 1) $ and for $ x $ bounded by the opening(s), we have
\begin{subequations} \label{eq11}
\allowdisplaybreaks
\addtolength{\jot}{0.5em}
\begin{align}
{\left. {{U_j}(x,z)} \right|_{z \to z_{bj}^ + }} & = {\left. {{U_{j + 1}}(x,z)} \right|_{z \to z_{bj}^ - }},  \label{eq11a} \\
{\left. {{\partial _z}{U_j}(x,z)} \right|_{z \to z_{bj}^ + }} & = {\left. {{\partial _z}{U_{j + 1}}(x,z)} \right|_{z \to z_{bj}^ - }}. \label{eq11b}
\end{align}
\end{subequations}
We will apply these boundary conditions to find $U_{j}(x,z)$ later.

\section{Application of Green's theorem}
\label{s3}
We assume that the diffracted field in the incident and transmission regions vanish at infinity; in this case, we can express $U_{j}(x, z)$ in region $j$ via the Green's theorem:
\begin{equation} \label{eq12}
{U_j}(x,z) = \int_{opening} {({G_j}{\partial _n}{U_j} - {U_j}{\partial _n}{G_j})ds},
\end{equation}
where $G_{j}$ is the Green's function for the region.

In contrast to the mode expansion methods that usually describe the mode projection \cite{Bravo, Martin, Garcia2, Villate2} as the Green's function \cite{Bravo, Martin, Garcia2, Villate2}, this method employs the standard point source radiation \cite{Goodman} to find the corresponding Green's function for our formulation. For an unbounded 2D system, the Green's function $ G $ is the magnetic field of the radiated cylindrical wave from a point source. Suppose that $(x', z')$ is the source coordinate to represent its location and $(x, z)$ is the observation coordinate, then the function should satisfy
\begin{equation} \label{eq13}
({\nabla ^2} + {k_0^2})G =  - \delta (x - x',z - z').
\end{equation}
We have the special function solution to the above equation \cite{Kreyszig}:
\begin{equation} \label{eq14}
G(x,z;x',z') = \frac{i}{4}H_0^{(1)}(k_0 R'),
\end{equation}
where $ H_{0}^{(1)}(k_0 R') $ is the Hankel function of the first kind and
\begin{equation} \label{eq15}
R' = {[{(x - x')^2} + {(z - z')^2}]^{1/2}}.
\end{equation}
The function $ H_{0}^{(1)}(k_0 R') $ is complex and its magnitude decays as $ (R')^{-1/2} $ when $ R' \to \infty$.

\begin{figure}[htbp]
\centering\includegraphics{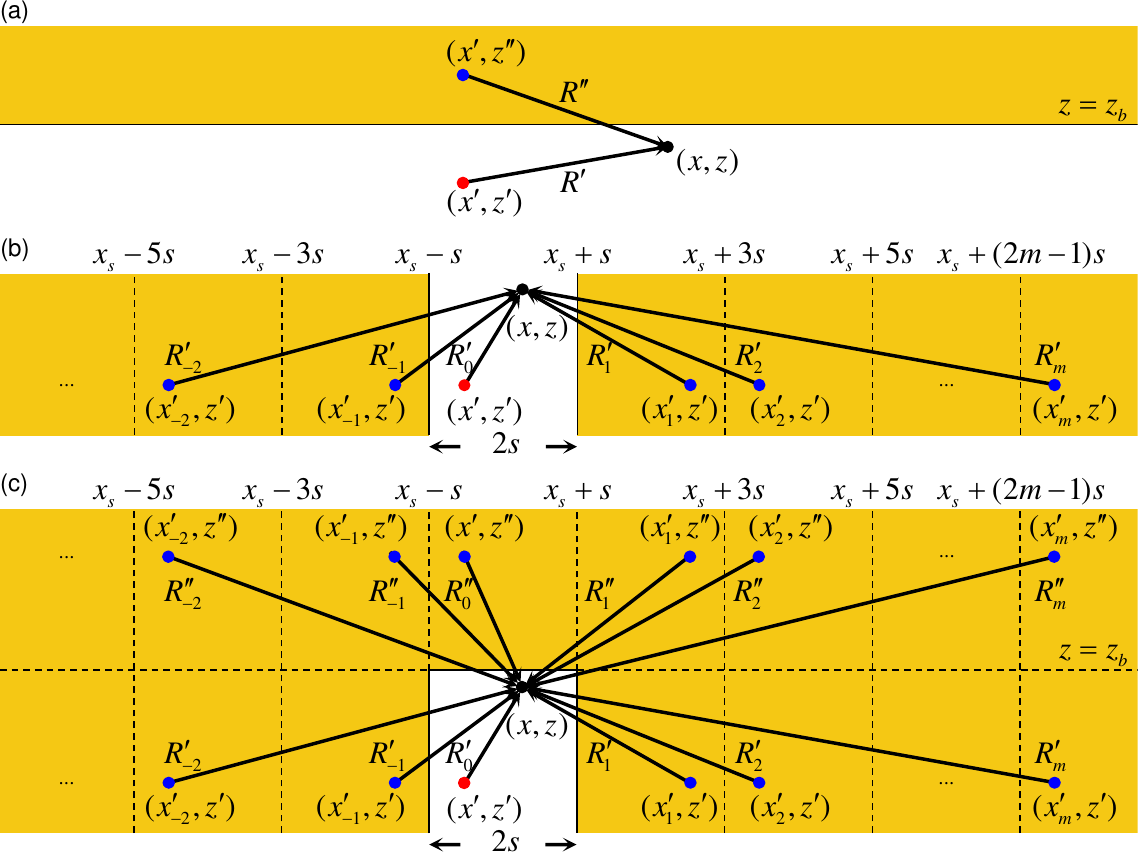}
\caption{Schematic of the influence on the observation point (black dot) from a real source (red dot) and its corresponding imaginary source(s) [blue dot(s)] under different types of boundaries: (a) a horizontal boundary at $z = z_{b}$, (b) two vertical boundaries at $x = x_{s} - s$ and $x = x_{s} + s$, and (c) the combination of the previous two types.}
\label{Fig02}
\end{figure}

In our wave interaction problem, the system is always bounded and the reflection from the point source radiation should be taken into account. Since $G$ represents the magnetic field, then similar to Eq. (\ref{eq03}), each boundary should let $ \partial_{n}G $ vanish, i.e.,
\begin{equation} \label{eq16}
\partial_{n} G = 0 \quad {\text{at the conductor surface}}.
\end{equation}
We organize three types of boundaries in Fig. \ref{Fig02} that will appear in the configuration of interest. To satisfy the condition, the method of images \cite{Morse} is employed. The radiation from the imaginary source(s) therefore describes the reflection. Then, the superposition of the radiation from all sources yields the corresponding Green's function.

For the first type in Fig. \ref{Fig02}(a), there exists only one boundary at $ z = z_{b} $. The corresponding imaginary source then is placed at the point $ (x', z'') $, where $ z'' = 2z_{b} - z' $. The superposition of the radiation from the real and imaginary sources yields the Green's function of the first type:
\begin{multline} \label{eq17}
\qquad \qquad \qquad \quad {G^{(1)}}(x,z;x',z';{z_b}) = \frac{i}{4}[H_0^{(1)}(k_0 R') + H_0^{(1)}(k_0 R'')]  \\ {\text{for}} -\infty<x,x'<\infty {\text{ and }} (z' - z_{b})(z - z_{b}) > 0,
\end{multline} 
where
\begin{equation} \label{eq18}
R'' = {[{(x - x')^2} + {(z + z' - 2{z_b})^2}]^{1/2}}.
\end{equation}
We consider the Green's function of this type as the description of the cylindrical wave mode in free space with the presence of the structure.

The second type in Fig. \ref{Fig02}(b) is for a space bounded by two vertical boundaries separated at a distance $2s$ and centered at $x = x_{s}$. The point sources at $(x'_{-1}, z')$ and $(x'_{1}, z')$ are placed to satisfy Eq. (\ref{eq16}) for the boundaries at $x = x_{s} - s$ and $x = x_{s} + s$, respectively. However, in this case, the influence from the left point source to the right boundary will not satisfy Eq. (\ref{eq16}), as well as the right point source to the left boundary. Thus, additional point sources at $(x'_{2}, z')$ and $(x'_{-2}, z')$ should be generated to seek the balance. Again, the added point sources can cause extra influence to the opposite boundaries, and further point sources should be placed. The process repeats infinitely to approach the final boundary conditions. Therefore, we obtain the Green's function of the second type:
\begin{multline} \label{eq19}
\qquad \qquad \qquad \qquad {G^{(2)}}(x,z;x',z';{x_x},s) = \frac{i}{4}\sum\limits_{m =  - \infty }^\infty  {H_0^{(1)}(k_0 {{R'_m}})}  \\ {\text{for }}{x_s} - s < x,x' < {x_s} + s,
\end{multline}
where
\begin{equation} \label{eq20}
{R'_m} = {[{(x - {x'_m})^2} + {(z - z')^2}]^{1/2}},
\end{equation}
and
\begin{equation} \label{eq21}
{x'_m} = (1 - q){x_s} + 2ms + q{x'}
\end{equation}
for $q = -1$ if $m$ is odd, or $q = 1$ if $m$ is otherwise. According to Ref. \cite{Morse}, the analytical solution to Eq. (\ref{eq19}) is
\begin{align} \label{eq22}
{G^{(2)}}(x,z;x',z';{x_s},s) & = \frac{i}{{4s{\gamma _0}}}{e^{i({\gamma _0}|z - z'|)}} \nonumber \\ 
& + \frac{i}{{2s}}\sum\limits_{m = 1}^\infty  {\frac{1}{{{\gamma _m}}}\cos [\frac{{m\pi (x - {x_s} + s)}}
{{2s}}]\cos [\frac{{m\pi (x' - {x_s} + s)}}{{2s}}]{e^{i({\gamma _m}|z - z'|)}}},
\end{align}
where
\begin{equation} \label{eq23}
{\gamma _m} = {[{k_0^2} - {(\frac{{m\pi }}{{2s}})^2}]^{1/2}}.
\end{equation}
By observing Eq. (\ref{eq22}), we can find that at least one propagating wave mode can exist in the bounded space, no matter how narrow it is. For the high-order modes, the field oscillates in the $x$ direction. The equation requires $2s$ greater than $(m\pi/k_0)$ to let the $m$th-order mode propagate. The observation meets the elementary waveguide calculation \cite{Kim1}. Therefore, Eq. (\ref{eq22}) describes the waveguide modes inside a slit. In addition, these modes are symmetric with respect to $z'$.

In this article, we propose a third type of space bounded as shown Fig. \ref{Fig02}(c). It is the combination of the first and the second types: there is one horizontal boundary at $z = z_{b}$ and two vertical boundaries separated at a distance $2s$ and centered at $x = x_{s}$. Besides the infinite imaginary sources at the same level of the real source, there are also infinite imaginary sources located at the opposite positions, corresponding to the horizontal boundary. As a result, we have the Green's function of third type:
\begin{align} \label{eq24}
{G^{(3)}}(x,z;x',z';{x_s},s,{z_b}) & = \frac{i}
{4}\sum\limits_{m =  - \infty }^\infty  {[H_0^{(1)}(k{{R'_m}}) + H_0^{(1)}(k{{R''_m}})]} \nonumber \\
& = {G^{(2)}}(x,z;x',z';{x_s},s) + {G^{(2)}}(x,z;x',2{z_b} - z';{x_s},s),
\end{align}
where
\begin{equation} \label{eq25}
{R''_m} = {[{(x - {x'_m})^2} + {(z + z' - 2{z_b})^2}]^{1/2}}.
\end{equation}
With the Green's function of this type, we will be able to describe the waveguide modes inside a groove. In contrast to the second type, these modes become asymmetric due to the horizontal boundary.

\section{Single slit}
\label{s4}
\begin{figure}[htbp]
\centering\includegraphics{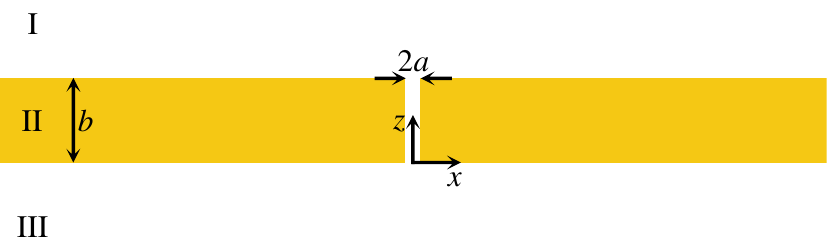}
\caption{Schematic of a subwavelength slit of width $2a$ and thickness $b$.}
\label{Fig03}
\end{figure}

A slit of width $2a$ and thickness $b$ is shown in Fig. \ref{Fig03}. The system is divided into three regions where the coordinate origin is located at the center of the slit.

As the entrance opening at $z = b$ and the exit opening at $z = 0$ are assumed to be short circuited \cite{Butler, Harrington, Chang1, Haddadpour}, the virtual boundaries result in the Green's functions at the interfaces that should satisfy
\begin{subequations} \label{eq26}
\addtolength{\jot}{0.5em}
\begin{align}
{\left. {{\partial _{z'}}{G_1}(x,z;x',z')} \right|_{z' \to {b^{+} }}} = 0, \label{eq26a} \\
{\left. {{\partial _{z'}}{G_3}(x,z;x',z')} \right|_{z' \to {0^{-} }}} = 0. \label{eq26b}
\end{align}
\end{subequations}
Hence, they are of the first type and can be written as
\begin{subequations} \label{eq27}
\addtolength{\jot}{0.5em}
\begin{align}
{G_1}(x,z;x',z') = {G^{(1)}}(x,z;x',z';b), \label{eq27a} \\
{G_3}(x,z;x',z') = {G^{(1)}}(x,z;x',z';0). \label{eq27b}
\end{align}
\end{subequations}
In region II, since we have assumed the virtual boundaries at the entrance and exit openings, $G_{2}$ in this region becomes unbounded in the $ z $ direction and is thus the Green's function of the second type:
\begin{equation} \label{eq28}
{G_2}(x,z;x',z') = {G^{(2)}}(x,z;x',z';0,a).
\end{equation}

With the explicit Green's functions and the boundary condition in Eq. (\ref{eq11b}), the integral representation for $U_j$ given in Eq. (\ref{eq12}) in each region can be expressed as
\begin{subequations} \label{eq29}
\allowdisplaybreaks
\begin{align}
{U^d}(x,z) =  & - \int_{ - a}^a {{G_1}(x,z;x',b){DU_{b}(x')}dx'}, \label{eq29a} \\
{U_3}(x,z) = & \int_{ - a}^a {{G_3}(x,z;x',0){DU_{0}(x')}dx'}, \label{eq29b} \\
{U_2}(x,z) = & - \int_{ - a}^a {[{G_2}(x,z;x',0)D{U_0}(x')} - {{U_0}(x'){{\left. {{\partial _{z'}}{G_2}(x,z;x',z')} \right|}_{z' \to {0^{+} }}}]dx'} \nonumber \\
& + \int_{ - a}^a {[{G_2}(x,z;x',b)D{U_b}(x')} - {{U_b}(x'){{\left. {{\partial _{z'}}{G_2}(x,z;x',z')} \right|}_{z' \to {b^{-} }}}]dx'}, \label{eq29c}
\end{align}
\end{subequations}
where we define the boundary fields and their normal derivatives at the entrance opening $z = b$ and the exit opening $z = 0$ as
\begin{subequations} \label{eq30}
\addtolength{\jot}{0.5em}
\allowdisplaybreaks
\begin{align}
{U_b}(x)  & \equiv {U_2}(x,b), \label{eq30a} \\
D{U_b}(x) & \equiv {\left. {{\partial _z}{U_2}(x,z)} \right|_{z \to {b^ - }}}, \label{eq30b} \\
{U_0}(x)  & \equiv {U_2}(x,0), \label{eq30c} \\
D{U_0}(x) & \equiv {\left. {{\partial _z}{U_2}(x,z)} \right|_{z \to {0^ + }}}. \label{eq30d}
\end{align}
\end{subequations} 
Then, we use the boundary condition in Eq. (\ref{eq11a}) for $z = b$ and $z = 0$ to yield a set of simultaneous integral equations:
\begin{subequations} \label{eq31}
\addtolength{\jot}{0.5em}
\allowdisplaybreaks
\begin{align} 
2U_b^i(x) - {U_b}(x) = & \int_{ - a}^a {{G_1}(x,b;x',b)D{U_b}(x')dx'}, \label{eq31a} \\
{U_0}(x) = & \int_{ - a}^a {{G_3}(x,0;x',0)D{U_0}(x')dx'}, \label{eq31b} \\
{U_b}(x) =  & - \int_{ - a}^a {[{G_2}(x,b;x',0)D{U_0}(x') - {U_0}(x'){{\left. {{\partial _{z'}}{G_2}(x,b;x',z')} \right|}_{z' \to {0^{+} }}}]dx'} \nonumber \\
& + \int_{ - a}^a {[{G_2}(x,b;x',b)D{U_b}(x') - {U_b}(x'){{\left. {{\partial _{z'}}{G_2}(x,b;x',z')} \right|}_{z' \to {b^{-} }}}]dx'}, \label{eq31c} \\
{U_0}(x) =  & + \int_{ - a}^a {[{G_2}(x,0;x',b)D{U_b}(x') - {U_b}(x'){{\left. {{\partial _{z'}}{G_2}(x,0;x',z')} \right|}_{z' \to {b^{-} }}}]dx'} \nonumber \\
& - \int_{ - a}^a {[{G_2}(x,0;x',0)D{U_0}(x') - {U_0}(x'){{\left. {{\partial _{z'}}{G_2}(x,0;x',z')} \right|}_{z' \to {0^{+} }}}]dx'}, \label{eq31d} 
\end{align}
\end{subequations}
where
\begin{equation} \label{eq32d} 
U_b^i(x) = \exp (ik_0 b).
\end{equation}
The solution to the unknowns $U_{b}(x)$, $DU_{b}(x)$, $U_{0}(x)$, and $DU_{0}(x)$ can be found in \ref{sa}, where a decritization method for the integrals is employed \cite{Betzig, Neerhoff}. Once the unknowns are found, they can be substituted into Eq. (\ref{eq29}) for $U_j$ in each region. Then, the electric field $\mathbf{E}$ is obtained from Eq. (\ref{eq02}).

\begin{figure}[htbp]
\centering\includegraphics{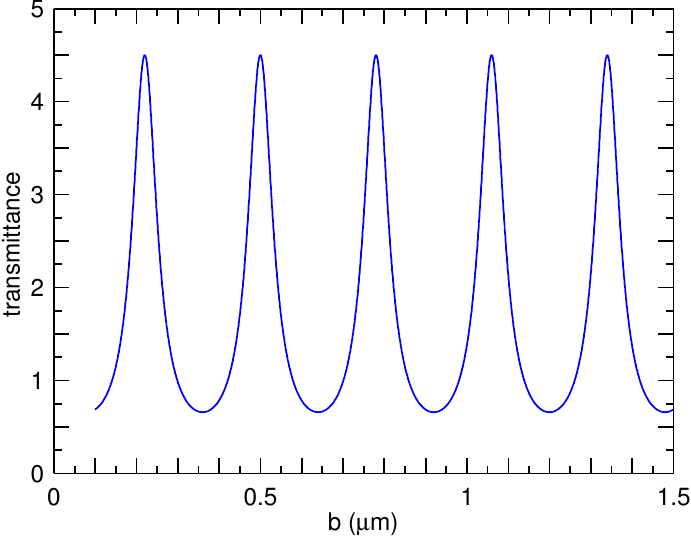}
\caption{Transmittance $T_s$ as a function of the film thickness $b$ for $\lambda_0 = 560$ nm and $2a = 40$ nm.}
\label{Fig04}
\end{figure}

To verify the analytical method, we assume $\lambda_0 = 560$ nm and $2a = 40$ nm; the number of the subintervals for resolving the boundary fields at the slit openings is $N = 8$. With the solutions yielded, the transmittance of the incident light through the slit is 
\begin{equation} \label{eq33}
T_s = \frac{1}{2a}\int_{-a}^{a} \frac{1}{2} \mathrm{Re} \{-E_x(x,0)U_0^*(x)\} dx.
\end{equation}
The result is shown in Fig. \ref{Fig04}. We obtain that $T_s$ varies periodically with the film thickness $b$ and the period is equal to $\lambda_0/2$. It indicates the Fabry-P\'{e}rot-like resonance, as revealed in the exteinsive studies \cite{Takakura, Chang3, Hong1, Li4, Bravo, Gordon1}. Moreover, $T_s$ at the peak is in good agreement with the theoretical prediction $(k_0 a)^{-1} = 4.46$ \cite{Betzig, Harrington}. Therefore, we establish the validity of our solution for the configuration of a single slit. 

\section{Slit patterned with grooves}
\label{s5}
\begin{figure}[htbp]
\centering\includegraphics{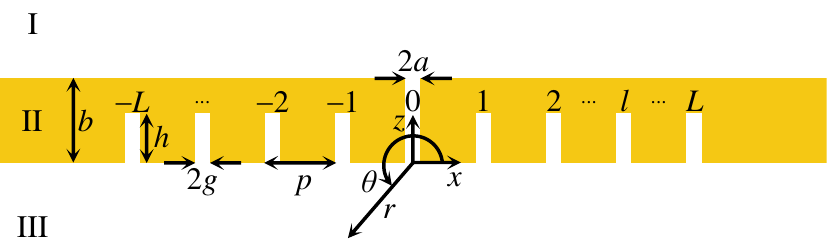}
\caption{Schematic of a subwavelength slit of width $2a$ and thickness $b$ that is patterned with $L$ pairs of grooves of width $2g$ and height $h$; the separation distance is $p$.}
\label{Fig05}
\end{figure}

This section shows the analytical method for a subwavelength slit patterned with grooves, as shown in Fig. \ref{Fig05}. There are $L$ pairs of grooves of width $2g$ and height $h$ patterned at the both sides of the slit exit while the separation distance is $p$.

Similar to the single slit case, there are three regions for the analysis. The Green's functions $G_{1}$ and $G_{3}$ share the same formulation as Eq. (\ref{eq27}). For the slit area in region II, $G_{2}$ remains the Green's function of the second type; for the groove areas, however, $G_{2}$ belongs to the third type because of the boundary at $z = h$. Since the Green's function is independent for the slit and each groove, we can use the rectangular function to represent the spatial characteristic of $G_{2}$:
\begin{align} \label{eq34}
{G_2}(x,z;x',z') & = {\text{rect}}(\frac{x}{{2a}}){\text{rect}}(\frac{{x'}}{{2a}}){G^{(2)}}(x,z;x',z';0,a) \nonumber \\
& + \sum\limits_{l =  - L,l \ne 0}^L {{\text{rect}}(\frac{{x - lp}}{{2g}}){\text{rect}}(\frac{{x' - lp}}{{2g}}){G^{(3)}}(x,z;x',z';lp,g,h).}
\end{align}

Similar to the case of the single slit, the integral representation for $U_j$ given in Eq. (\ref{eq12}) in each region becomes
\begin{subequations} \label{eq35}
\addtolength{\jot}{0.5em}
\allowdisplaybreaks
\begin{align}
{U^d}(x,z) =  & - \int_{ - a}^a {{G_1}(x,z;x',b){DU_{b}(x')}dx'}, \label{eq35a} \\ 
{U_3}(x,z) = & \int_{ -Lp-g}^{Lp+g} {{G_3}(x,z;x',0){DU_{0}(x')}dx'}, \label{eq35b} \\
{U_2}(x,z) = & - \int_{ - Lp - g}^{Lp + g} {[{G_2}(x,z;x',0)D{U_0}(x')}{- {U_0}(x'){{\left. {{\partial _{z'}}{G_2}(x,z;x',z')} \right|}_{z' \to {0^{+} }}}]dx'} \nonumber \\
& + \int_{ - a}^a {[{G_2}(x,z;x',b)D{U_b}(x')}{- {U_b}(x'){{\left. {{\partial _{z'}}{G_2}(x,z;x',z')} \right|}_{z' \to {b^{-} }}}]dx'}. \label{eq35c}
\end{align}
\end{subequations}
The boundary condition in Eq. (\ref{eq11a}) then yields a set of simultaneous integral equations:
\begin{subequations} \label{eq36}
\addtolength{\jot}{0.5em}
\allowdisplaybreaks
\begin{align} 
2U_b^i(x) - {U_b}(x) = & \int_{ - a}^a {{G_1}(x,b;x',b)D{U_b}(x')dx'}, \label{eq36a} \\
{U_0}(x) = & \int_{-Lp-g}^{Lp+g} {{G_3}(x,0;x',0)D{U_0}(x')dx'}, \label{eq36b} \\
{U_b}(x) =  & - \int_{ - a}^a {[{G_2}(x,b;x',0)D{U_0}(x') - {U_0}(x'){{\left. {{\partial _{z'}}{G_2}(x,b;x',z')} \right|}_{z' \to {0^{+} }}}]dx'} \nonumber \\
& + \int_{ - a}^a {[{G_2}(x,b;x',b)D{U_b}(x') - {U_b}(x'){{\left. {{\partial _{z'}}{G_2}(x,b;x',z')} \right|}_{z' \to {b^{-} }}}]dx'}, \label{eq36c} \\
{U_0}(x) =  & + \int_{-a}^{a} {[{G_2}(x,0;x',b)D{U_b}(x') - {U_b}(x'){{\left. {{\partial _{z'}}{G_2}(x,0;x',z')} \right|}_{z' \to {b^{-} }}}]dx'} \nonumber \\
& - \int_{-Lp-g}^{Lp+g} {[{G_2}(x,0;x',0)D{U_0}(x') - {U_0}(x'){{\left. {{\partial _{z'}}{G_2}(x,0;x',z')} \right|}_{z' \to {0^{+}}}}]dx'}. \label{eq36d} 
\end{align}
\end{subequations}
Similar to the solution of the single-slit configuration, we are able to find the unknowns and the EM field in each region.

\begin{figure}[htbp]
\centering\includegraphics{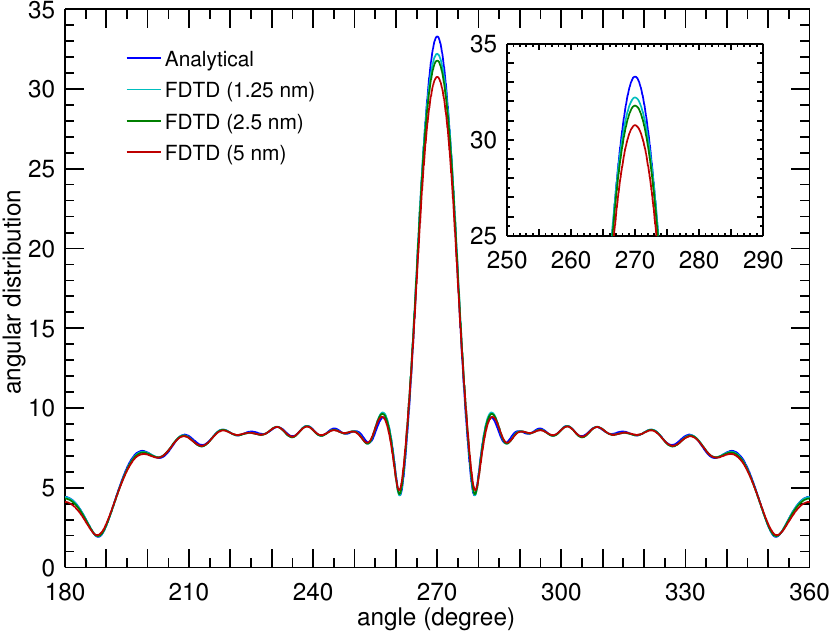}
\caption{Angular distribution from the analytical method (blue curve) and from the FDTD simulation when the cell size is $5$ nm (red curve), $2.5$ nm (green curve), and $1.25$ nm (cyan curve). Inset: enlarged view of the angular distributions near the angle $\theta = 270^\circ$.}
\label{Fig06}
\end{figure}

Suppose $\lambda_{0} = 560$ nm, $2a = 2g = 40$ nm, $b = 250$ nm, $h = 100$ nm, $p = 500$ nm, and $L = 10$; the numbers of the subintervals for the slit and the groove openings are both $N = 8$. We define the angular distribution of the field in region III as
\begin{equation} \label{eq37}
f(\theta) = {(\pi r)^{1/2}}|{U_3}(r,{\theta})|,
\end{equation}
where $r = (x^2 + z^2)^{1/2}$ and $\theta = \tan^{-1}(z/x)$. The resultant $f(\theta)$ is shown in Fig. \ref{Fig06} for $r = 20$ $\mu$m. We obtain a light beam in free space that is consistent with the previous study \cite{Hong2, Martin}. 

To further verify the result, we perform the FDTD simulation with the cell sizes of $5$ nm, $2.5$ nm, and $1.25$ nm, respectively, and show their angular distributions \cite{Hong2} in Fig. \ref{Fig06}. The distribution from the analytical method is well agreed with those from the simulation.  While the peak value from the analytical method is $33.3$, those from the simulation with the cell sizes of $5$ nm, $2.5$ nm, and $1.25$ nm are $ 30.8 $, $ 31.8 $, and $ 32.2 $, respectively. Therefore, as the cell size reduces, the distribution approaches to the analytical result. We estimate that, as the cell size approaches to zero, the peak value will approach to $32.7$, which is only $1.80\%$ different from the analytical result. We thus verify the accuracy of the method for the configuration.

\section{Double slit}
\label{s6}
\begin{figure}[htbp]
\centering\includegraphics{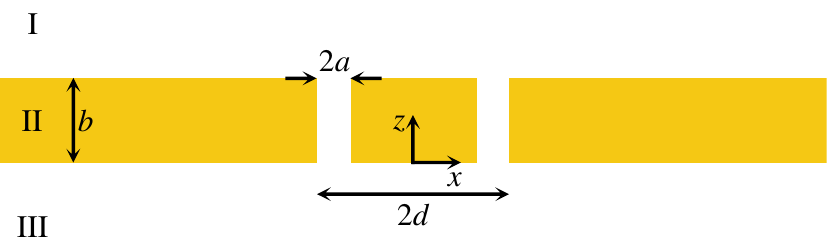}
\caption{Schematic of a double slit of width $2a$ and thickness $b$; the edge-to-edge distance is $2d$.}
\label{Fig07}
\end{figure}

In this configuration, a double slit of width $2a$ and thickness $b$ are located at both sides of a coordinate origin and the edge-to-edge distance is $ 2d $, as shown in Fig. \ref{Fig07}.

In region I and III, the Green's functions $G_{1}$ and $G_{3}$ share the same form as Eq. (\ref{eq27}). In region II, the formulation of $G_{2}$ is similar to the single-slit configuration but responsible to the field inside the two slits. To represent the spatial characteristic of $G_{2}$, the rectangular function is used:
\begin{align} \label{eq38}
{G_2}(x,z;x',z') & = {\text{rect}}(\frac{{x + d - a}}{{2a}}){\text{rect}}(\frac{{x' + d - a}}{{2a}}){G^{(2)}}(x,z;x',z'; - d + a,a) \nonumber \\
& + {\text{rect}}(\frac{{x - d + a}}{{2a}}){\text{rect}}(\frac{{x' - d + a}}{{2a}}){G^{(2)}}(x,z;x',z';d - a,a).
\end{align}

The integral representation for $U_j$ given in Eq. (\ref{eq12}) in each region becomes
\begin{subequations} \label{eq39}
\allowdisplaybreaks
\addtolength{\jot}{0.5em}
\begin{align}
{U^d}(x,z) =  & - \int_{ - d}^d {{G_1}(x,z;x',b){DU_{b}(x')}dx'}, \label{eq39a} \\
{U_3}(x,z) = & \int_{ -d}^{d} {{G_3}(x,z;x',0){DU_{0}(x')}dx'}, \label{eq39b} \\
{U_2}(x,z) = & - \int_{-d}^{d} {[{G_2}(x,z;x',0)D{U_0}(x')}{- {U_0}(x'){{\left. {{\partial _{z'}}{G_2}(x,z;x',z')} \right|}_{z' \to {0^{+} }}}]dx'} \nonumber \\
& + \int_{-d}^d {[{G_2}(x,z;x',b)D{U_b}(x')}{- {U_b}(x'){{\left. {{\partial _{z'}}{G_2}(x,z;x',z')} \right|}_{z' \to {b^{-} }}}]dx'}. \label{eq39c}
\end{align}
\end{subequations}
The boundary conditions given in Eq. (\ref{eq11a}) yields a set of simultaneous integral equations:
\begin{subequations} \label{eq40}
\allowdisplaybreaks
\addtolength{\jot}{0.5em}
\begin{align} 
2U_b^i(x) - {U_b}(x) = & \int_{ - d}^d {{G_1}(x,b;x',b)D{U_b}(x')dx'}, \label{eq40a} \\
{U_0}(x) = & \int_{-d}^{d} {{G_3}(x,0;x',0)D{U_0}(x')dx'}, \label{eq40b} \\
{U_b}(x) =  & - \int_{ - d}^d {[{G_2}(x,b;x',0)D{U_0}(x') - {U_0}(x'){{\left. {{\partial _{z'}}{G_2}(x,b;x',z')} \right|}_{z' \to {0^ + }}}]dx'} \nonumber \\
& + \int_{ - d}^d {[{G_2}(x,b;x',b)D{U_b}(x') - {U_b}(x'){{\left. {{\partial _{z'}}{G_2}(x,b;x',z')} \right|}_{z' \to {b^ - }}}]dx'}, \label{eq40c} \\
{U_0}(x) =  & + \int_{-d}^{d} {[{G_2}(x,0;x',b)D{U_b}(x') - {U_b}(x'){{\left. {{\partial _{z'}}{G_2}(x,0;x',z')} \right|}_{z' \to {b^ - }}}]dx'} \nonumber \\
& - \int_{ - d}^d {[{G_2}(x,0;x',0)D{U_0}(x') - {U_0}(x'){{\left. {{\partial _{z'}}{G_2}(x,0;x',z')} \right|}_{z' \to {0^ + }}}]dx'}. \label{eq40d} 
\end{align}
\end{subequations}
The unknowns can also be found with the solution similar to the single-slit configuration, so can the EM field in each region.

\begin{figure}[htbp]
\centering\includegraphics[width=6.53cm]{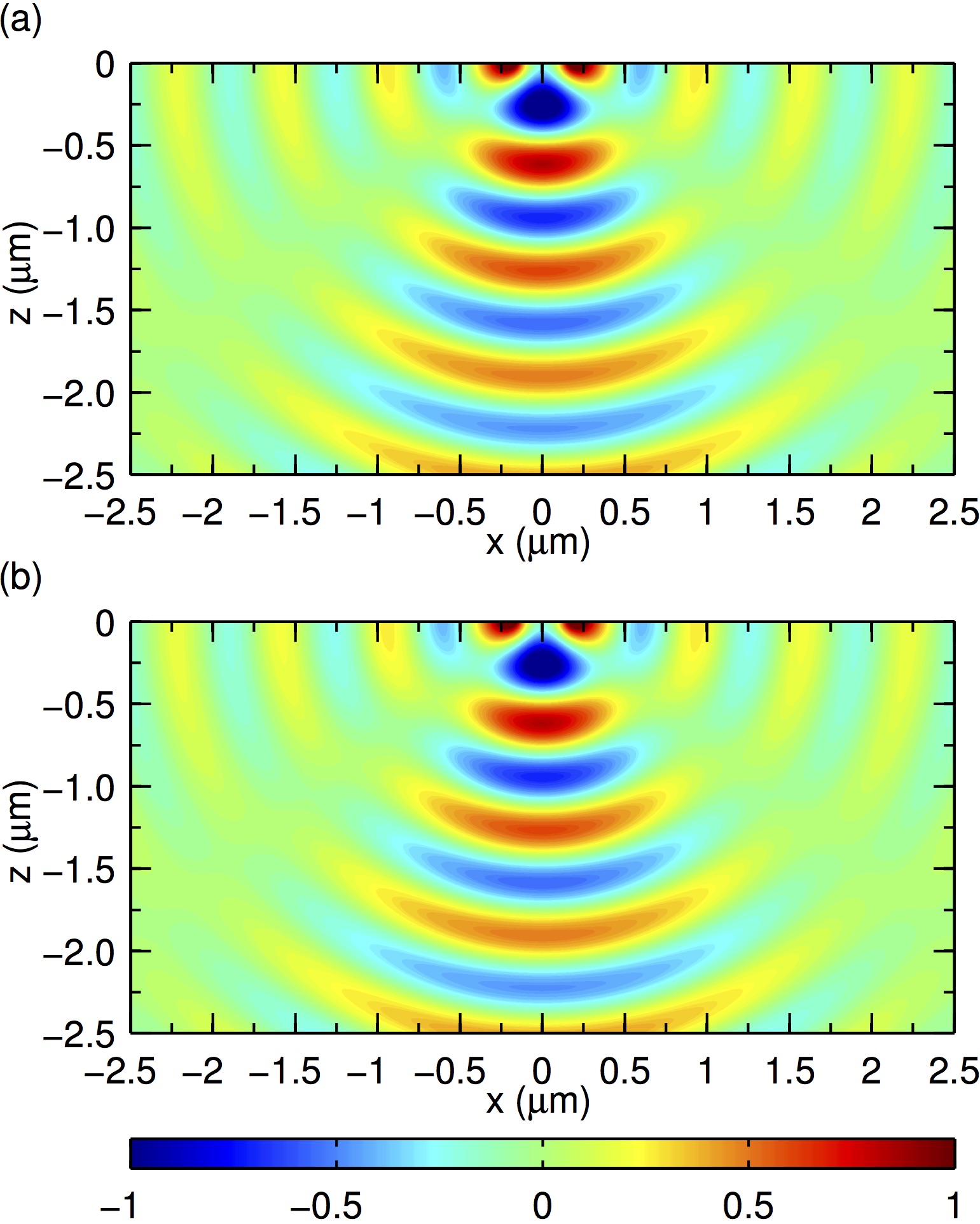}
\caption{(a) Spatial field distribution of the resultant Re$\{U_{3}(x,z)\exp(-i\omega t)\}$ from the analytical method and (b) snapshot of the magnetic field from the FDTD simulation.}
\label{Fig08}
\end{figure}

In the configuration, we let $\lambda_{0} = 633$ nm, $2a = 80$ nm, $b = 200$ nm, $2d = 480$ nm, and $N = 16$ for each opening of the slits. We show the snapshot of the field Re$\{U_{3}(x,z)\exp(-i\omega t)\}$ in Fig. \ref{Fig08}(a) at the temporal phase $\omega t = 0.95 \times 2\pi$; the temporal phase is chosen when the central field at the exit surface is about zero. To compare, we perform the FDTD simulation with the cell size of $5$ nm, and show the snapshot of the field distribution in Fig. \ref{Fig08}(b). The simulation time is chosen also for the central field to be about zero. We obtain that the profile and the magnitude of the field from both the analytical method and the simulation are in excellent agreement.

\begin{figure}[htbp]
\centering\includegraphics{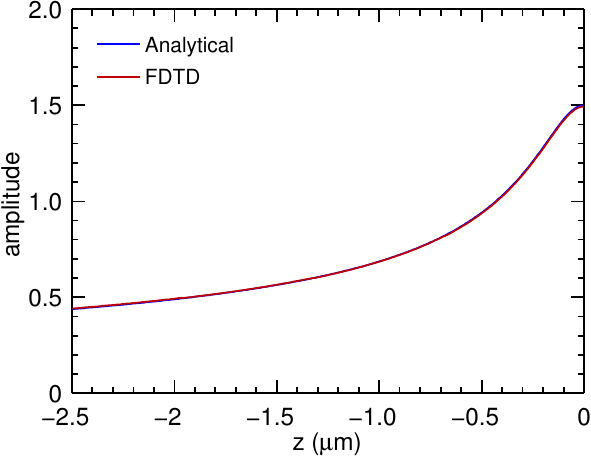}
\caption{Amplitude distributions along the vertical line $ x = 0 $ from the analytical method (blue curve) and from the FDTD simulation (red curve).}
\label{Fig09}
\end{figure}

To further confirm the results, we show in Fig. \ref{Fig09} the field amplitude along the vertical line $x = 0$ from the analytical method, as well as that from the FDTD simulation. The comparison evidences the accuracy of the analytical method.

\section{Indented double slit}
\label{s7}
\begin{figure}[htbp]
\centering\includegraphics{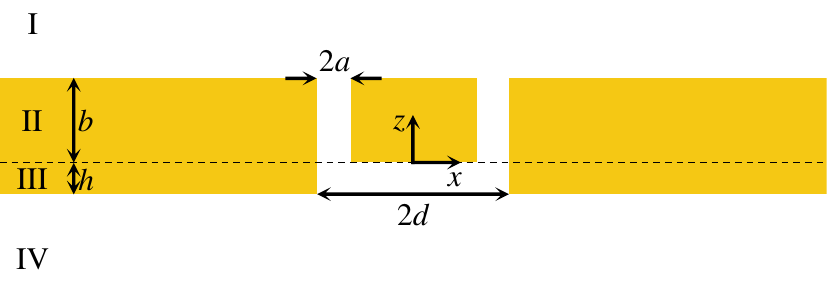}
\caption{Schematic of an indented double slit. In the configuration, the structure region is divided into two sub-regions composed of a double slit and a single slit, respectively. The width, thickness, and separation distance of the double slit are $2a$, $b$, and $2d$, respectively, while $h$ is the thickness of the single slit.}
\label{Fig10}
\end{figure}

The last configuration is an indented double slit, as shown in Fig. \ref{Fig10}; in the configuration, a certain amount of the central metal between the two slit is removed. From the aspect of the analytical method, the configuration can be considered as a double slit serially connected to a single slit of width equal to the separation distance. Therefore, it is a two-layered hybrid configuration. The structure region is divided into two sub-regions to represent the magnetic field in these two components. Similar to the double-slit case, $2a$, $b$, and $2d$ are denoted as the width, thickness, and separation distance of the double slit, respectively, while $h$ represents the thickness of the single slit.

The incident and transmission regions in the configuration are represented as regions I and IV, respectively. Their corresponding Green's functions are
\begin{subequations} \label{eq41}
\allowdisplaybreaks
\addtolength{\jot}{0.5em}
\begin{align}
G_{1}(x,z;x',z') & = G^{(1)}(x,z;x',z';b), \label{eq41a} \\
G_{4}(x,z;x',z') & = G^{(1)}(x,z;x',z';-h). \label{eq41b}
\end{align}
\end{subequations}
The Green's function in region II $ G_2(x,z;x',z') $ is as the same as that in the double-slit configuration. The Green's function in region III, however, is given by two different boundary conditions at the interfaces $z = 0$ and $z = -h$, respectively. At $z = 0$, a virtual boundary is assumed for the two openings of the double slit to be short circuited. Thus, the corresponding Green's function at the interface $ G_{23} $ should satisfy
\begin{equation} \label{eq42}
{\left. {{\partial _{z'}}{G_{23}}(x,z;x',z')} \right|_{z' \to {0^ - }}} = 0.
\end{equation}
Since the region is also bounded vertically at $|x| = d$, $ G_{23} $ is of the third type. Therefore, we give the Green's function for the interface
\begin{equation} \label{eq43}
{G_{23}}(x,z;x',z') = {G^{(3)}}(x,z;x',z';0,d,0).
\end{equation}
At $z = -h$, the virtual boundary assumption for the double-slit openings results in the unbounded condition in the $z$ direction with respect to the interface. The corresponding Green's function $ G_{34} $ is therefore of the second type, which gives
\begin{equation} \label{eq44}
{G_{34}}(x,z;x',z') = {G^{(2)}}(x,z;x',z';0,d).
\end{equation}

With the explicit Green's functions and the boundary condition in Eq. (\ref{eq11b}), the integral representation for $U_j$ given in Eq. (\ref{eq12}) in each region can be expressed as
\begin{subequations} \label{eq45}
\allowdisplaybreaks
\addtolength{\jot}{0.5em}
\begin{align}
{U^d}(x,z) = & - \int_{ - d}^d {{G_1}(x,z;x',b){DU_{b}(x')}dx'}, \label{eq45a} \\
{U_4}(x,z) = & \int_{ -d}^{d} {{G_4}(x,z;x',-h){DU_{h}(x')}dx'}, \label{eq45b} \\
{U_2}(x,z) = & - \int_{-d}^{d} {[{G_2}(x,z;x',0)D{U_0}(x')}{- {U_0}(x'){{\left. {{\partial _{z'}}{G_2}(x,z;x',z')} \right|}_{z' \to {0^ + }}}]dx'} \nonumber \label{eq45c} \\
& + \int_{-d}^d {[{G_2}(x,z;x',b)D{U_b}(x')}{- {U_b}(x'){{\left. {{\partial _{z'}}{G_2}(x,z;x',z')} \right|}_{z' \to {b^ - }}}]dx'}, \\
{U_3}(x,z) =  & - \int_{ - d}^d {[{G_{34}}(x,z;x', - h)D{U_h}(x')}{- {U_h}(x'){{\left. {{\partial _{z'}}{G_{34}}(x,z;x',z')} \right|}_{z' \to  - {h^ + }}}]dx'} \nonumber \label{eq45d} \\
& + \int_{ - d}^d {{G_{23}}(x,z;x',0)D{U_0}(x')dx'},
\end{align}
\end{subequations}
where, besides the boundary fields and their normal derivatives $U_b(x)$, $DU_b(x)$, $U_0(x)$, and $DU_0(x)$, we define those at the interface $z = -h$ as
\begin{subequations} \label{eq46}
\addtolength{\jot}{0.5em}
\allowdisplaybreaks
\begin{align}
{U_h}(x) & \equiv {U_3}(x, - h), \label{eq46a} \\
D{U_h}(x) & \equiv {\left. {{\partial _z}{U_3}(x,z)} \right|_{z \to  - {h^ + }}}. \label{eq46b} 
\end{align}
\end{subequations}
Then, we use the boundary condition in Eq. (\ref{eq11a}) for $z = b$, $z = 0$, and $z = -h$ to yield a set of simultaneous integral equations:
\begin{subequations} \label{eq47}
\allowdisplaybreaks
\addtolength{\jot}{0.5em}
\begin{align} 
2U_b^i(x) - {U_b}(x) = & \int_{ - d}^d {{G_1}(x,b;x',b)D{U_b}(x')dx'}, \label{eq47a} \\
{U_h}(x) = & \int_{ - d}^d {{G_4}(x, - h;x', - h)D{U_h}(x')dx'}, \label{eq47b} \\
{U_b}(x) =  & - \int_{ - d}^d {[{G_2}(x,b;x',0)D{U_0}(x') - {U_0}(x'){{\left. {{\partial _{z'}}{G_2}(x,b;x',z')} \right|}_{z' \to {0^ + }}}]dx'} \nonumber \\
& + \int_{ - d}^d {[{G_2}(x,b;x',b)D{U_b}(x') - {U_b}(x'){{\left. {{\partial _{z'}}{G_2}(x,b;x',z')} \right|}_{z' \to {b^ - }}}]dx'}, \label{eq47c} \\
{U_0}(x) = & + \int_{-d}^{d} {[{G_2}(x,0;x',b)D{U_b}(x') - {U_b}(x'){{\left. {{\partial _{z'}}{G_2}(x,0;x',z')} \right|}_{z' \to {b^ - }}}]dx'} \nonumber \\
& - \int_{ - d}^d {[{G_2}(x,0;x',0)D{U_0}(x') - {U_0}(x'){{\left. {{\partial _{z'}}{G_2}(x,0;x',z')} \right|}_{z' \to {0^ + }}}]dx'}, \label{eq47d} \\ 
{U_0}(x) = & - \int_{ - d}^d {[{G_{34}}(x,0;x', - h)D{U_h}(x')} {- {U_h}(x'){{\left. {{\partial _{z'}}{G_{34}}(x,0;x',z')} \right|}_{z' \to  - {h^ + }}}]dx'} \nonumber \\
&　+ \int_{ - d}^d {{G_{23}}(x,0;x',0)D{U_0}(x')]dx'},　\label{eq47e}　\\
{U_h}(x) = & + \int_{ - d}^d {{G_{23}}(x, - h;x',0)D{U_0}(x')dx'} \nonumber \\
& - \int_{ - d}^d {[{G_{34}}(x, - h;x', - h)D{U_h}(x')} {- {U_h}(x'){{\left. {{\partial _{z'}}{G_{34}}(x, - h;x',z')} \right|}_{z' \to  - {h^ + }}}]dx'}. \label{eq47f}
\end{align}
\end{subequations}
We obtain that there are six equations for the same number of the unknowns $U_{b}(x)$, $DU_{b}(x)$, $U_{0}(x)$, $DU_{0}(x)$, $U_{h}(x)$, and $DU_{h}(x)$. Therefore, they are readily to be solved with the desensitization method for the integrals similar to the single-slit configuration. Then, they can be substituted into Eq. (\ref{eq45}) for the field in each region.

\begin{figure}[htbp]
\centering\includegraphics[width=2.57in]{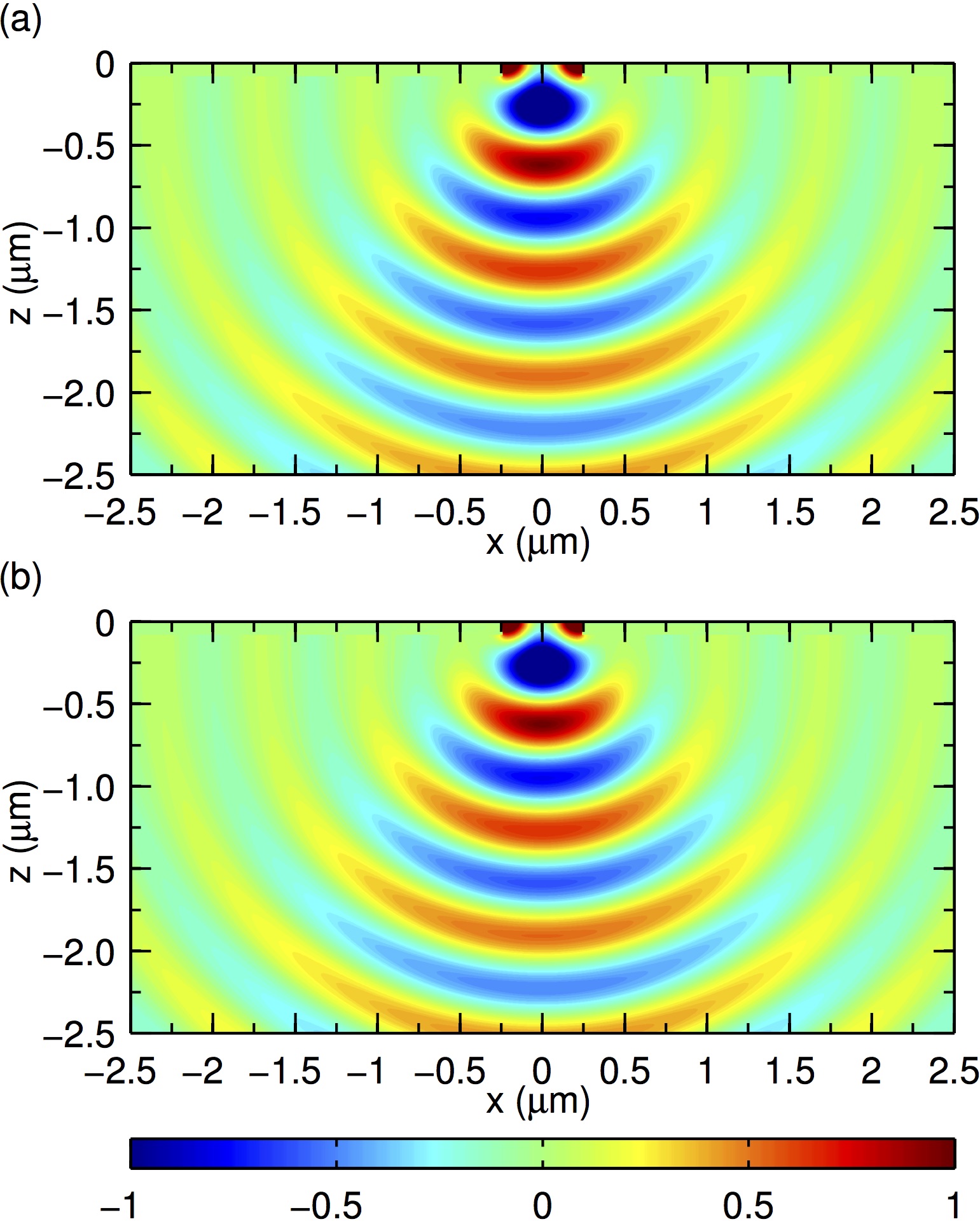} 
\caption{(a) Spatial field distribution of the resultant Re$\{U_{3}(x,z)\exp(-i\omega t)\}$ and Re$\{U_{4}(x,z)\exp(-i\omega t)\}$ and (b) snapshot of simulated magnetic field from the FDTD simulation.}
\label{Fig11}
\end{figure}

We assume that $\lambda_{0} = 633$ nm, $2a = 80$ nm, $b = 200$ nm, $2d = 480$ nm, $h = 80$ nm, $N = 16$ for each opening of the double slit, and $N = 96$ for the opening of the single slit. We show the snapshots Re$\{U_{3}(x,z)\exp(-i\omega t)\}$ and Re$\{U_{4}(x,z)\exp(-i\omega t)\}$ together in Fig. \ref{Fig11}(a) at the temporal phase $\omega t = 0.98 \times 2\pi$; similar to the double-slit configuration, the temporal phase is chosen when the central field at the exit surface of the double slit is about zero. We perform the FDTD simulation with the cell size of $5$ nm, and show the snapshot of the field distribution in Fig. \ref{Fig11}(b). The simulation time is chosen also to let the central field be about zero. Both the field profile and the magnitude are in excellent agreement. We notice that, with the indented structure, the transmitted field is more focused than that in the double-slit configuration, as has been indicated in Refs. \cite{Chen1, Chen2}.

\begin{figure}[htbp]
\centering\includegraphics{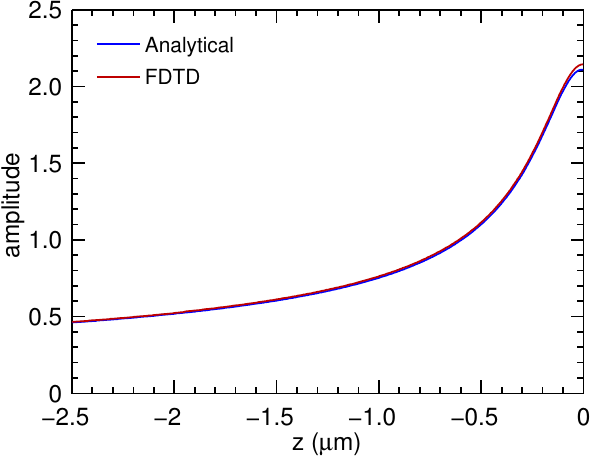}
\caption{Amplitude distributions along the vertical line $ x = 0 $ from the analytical method (blue curve) and from the FDTD simulation (red curve).}
\label{Fig12}
\end{figure}

We show in Fig. \ref{Fig12} the field amplitude along the vertical line $x = 0$ from the analytical method, as well as that from the FDTD simulation with the cell size of $5$ nm. The results are also in excellent agreement. The analytical method shows the accuracy for the hybrid configuration. Interestingly, it is noted that the field amplitude at $z = 0$ is increased by $40.5\%$, as compared with that in the double-slit configuration. Therefore, this configuration is indeed capable of enhancing the transmission significantly to have the academic value and the application potential.

\section{Discussions}
\label{s8}
In the previous sections, we showed that the rigorous formulation of the field in each region based on the Green's theorem is capable of providing accurate solutions to the problem of the light interaction with the various configurations of subwavelength structures. In this section, we discuss the applications of this method for the further physical interpretation and study.

In the single-slit configuration, since the width is much smaller than the wavelength, we can assume that the boundary fields $U_b(x)$ and $U_0(x)$, and their normal derivatives $DU_b(x)$ and $DU_0(x)$, are constant \cite{Martin, Garcia2}. We define the integral of the Green's functions in Eqs. (\ref{eq31a}) and (\ref{eq31b}) as
\begin{equation} \label{eq48}
I = \frac{i}{2} \int_{-a}^{a}{H_0^{(1)}(k_0|x|)dx}.
\end{equation}
Moreover, the $m \geq 1$ high-order terms of $G_2$ are negligible. From Eq. (\ref{eq31}), we then obtain the simplified simultaneous equations:
\begin{subequations} \label{eq49}
\allowdisplaybreaks
\addtolength{\jot}{0.5em}
\begin{align}
2U_b^i - U_b = & DU_bI, \label{eq49a} \\
U_0 = & DU_0I, \label{eq49b} \\
U_b = &-\frac{i}{2k_0}e^{ik_0 b}DU_0 + \frac{1}{2}e^{ik_0 b}U_0 + \frac{i}{2k_0}DU_b + \frac{1}{2}U_b, \label{eq49c} \\
U_0 = &-\frac{i}{2k_0}DU_0 + \frac{1}{2}e^{ik_0 b}U_b + \frac{i}{2k_0}e^{ik_0 b}DU_b + \frac{1}{2}U_0. \label{eq49d} 
\end{align}
\end{subequations}
The simplification of the formulation leads to the explicit analytical solutions to the unknowns:
\begin{subequations} \label{eq50}
\allowdisplaybreaks
\addtolength{\jot}{0.5em}
\begin{align}
U_0 = & \frac{4ik_0 I U_b^i e^{ik_0 b}}{(k_0 I + i)^2 - (k_0 I - i)^2 e^{i2k_0 b}}, \label{eq50a} \\
DU_0 = & \frac{4ik_0 U_b^i e^{ik_0 b}}{(k_0 I + i)^2 - (k_0 I - i)^2 e^{i2k_0 b}}, \label{eq50b} \\
U_b = & \frac{-2 U_b^i [(1 - e^{i2k_0 b}) - ik_0 I(1 + e^{i2k_0 b})]}{(k_0 I + i)^2 - (k_0 I - i)^2 e^{i2k_0 b}}, \label{eq50c} \\
DU_b = & \frac{2k_0 U_b^i [(1 - e^{i2k_0 b})k_0 I + i(1 + e^{i2k_0 b})]}{(k_0 I + i)^2 - (k_0 I - i)^2 e^{i2k_0 b}}. \label{eq50d}
\end{align}
\end{subequations}
According to Eq. (\ref{eq29c}), the simplification reduces the field inside the slit to 
\begin{equation} \label{eq51}
U_s(z) \equiv U_2(x,z) = \frac{i}{2k_0} [-DU_0 e^{ik_0 z} + U_0 \partial_{z'} e^{ik_0 (z - z')} |_{z' \to {0^ + }} + DU_b e^{ik_0 (b - z)} - U_b \partial_{z'} e^{ik_0 (z' - z)} |_{z' \to {b^ - }}].
\end{equation}
In the substitution of the solutions, we obtain
\begin{equation} \label{eq52}
U_s(z) = U_b^i e^{ik_0 b} t_0 \frac{e^{-ik_0 z} - r_0 e^{ik_0 z}}{1 - r_0^2 e^{i2k_0 b}},
\end{equation}
where $t_0 = (2i)/(k_0 I + i)$ and $r_0 = -(k_0 I - i) / (k_0 I + i)$. We find that the representation of the field inside the slit in fact describes the Fabry-P\'{e}rot-like resonance that results from the roundtrips of the propagating wave \cite{Chang3, Hong1, Li4}. Our analytical method yields the transmission coefficient $t_0$ for an incident plane wave through the $z = b$ opening and the reflection coefficient $r_0$ at both the openings $z = b$ and $z = 0$. Given the electric field $E_{xs}(z) = (-i/k_0)\partial_z U_s(z)$, the resultant transmittance $T_s = \mathrm{Re} \{-E_{xs}(z) U_s^*(z) \}$ as a function of the film thickness $b$ is consistent with those as shown in Fig. \ref{Fig04}.

For the configuration of the slit patterned with the grooves, we can also assume that the boundary fields and their normal derivatives at each groove opening are constant. In this case, we denote $U_0^l = U_0(x)$ and $DU_0^l = DU_0(x)$ when $|x - lp| < s$, where $s = a$ when $l = 0$ and $g$ otherwise. The integral of the Green's function $G_3$ in Eq. (\ref{eq36b}) for the radiated wave from the $k$th opening ($k = -L, \ldots L$) coupled to the $l$th opening is
\begin{equation} \label{eq53}
I_{l,k} = \frac{i}{2} \int_{kp-s}^{kp+s}{H_0^{(1)}(k_0|x-lp|)dx}.
\end{equation}
It should be noted that $I_{0,0} = I$ in the single-slit case. With the ignorance of the high-order terms of the Green's function $G_2$ for both the slit and the grooves, the simultaneous equations become from Eq. (\ref{eq36}) to:
\begin{subequations} \label{eq54}
\allowdisplaybreaks
\addtolength{\jot}{0.5em}
\begin{align}
2U_b^i - U_b = & DU_bI, \label{eq54a} \\
U_0^l = & \sum\limits_{k = -L}^{L} DU_0^k I_{l,k}, \label{eq54b} \\
U_b = &-\frac{i}{2k_0}e^{ik_0 b}DU_0^0 + \frac{1}{2}e^{ik_0 b}U_0^0 + \frac{i}{2k_0}DU_b + \frac{1}{2}U_b, \label{eq54c} \\
U_0^0 = &-\frac{i}{2k_0}DU_0^0 + \frac{1}{2}e^{ik_0 b}U_b + \frac{i}{2k_0}e^{ik_0 b}DU_b + \frac{1}{2}U_0^0, \label{eq54d} \\
U_0^l = &-\frac{i}{2k_0}(1 + e^{i2k_0 h})DU_0^l + \frac{1}{2}(1 + e^{i2k_0 h})U_0^l \quad {\text{for }} l \neq 0. \label{eq54e}
\end{align}
\end{subequations}
With the explicit equations, the unknowns are readily to be solved. The simplification reduces the field in region III to
\begin{equation} \label{eq56}
U_3(x,z) = \sum\limits_{l = -L}^{L}DU_0^l \int_{lp-s}^{lp+s}G_3(x,z;x',0)dx'.
\end{equation}
We obtain that, because the opening width is much smaller than the wavelength, the integrals for any $(x,z)$ away from the openings are constant. The resultant field $U_3(x,z)$ can be considered as the radiation from point sources at each opening \cite{Hong2}. Our method therefore provides the analytical solution to study the diffraction in the configuration of the groove-patterned subwavelength structures. 

As for the double-slit configuration, similar approximations can be made to simplify the simultaneous equations. In the configuration, it can be expected that the approximation will result in the similar Fabry-P\'{e}rot-like resonance inside the slits to the single-slit configuration, while the additional coupling between the slits is taken place \cite{Gordon2}. The approximation will therefore be helpful for us to analyze and model the double-slit coupling mechanism and transmission. 

In the configuration of the indented double slit, the analysis will not be simple because, in region III, the waveguides modes from the boundary fields at the $z = 0$ interface are asymmetric, while those at the $z = -h$ interface are symmetric. Besides, both of the high-order modes should be taken into account since the slit width is larger than half wavelength. Nevertheless, we can still take a peek of the interesting complex coupling mechanism between the interfaces in region III with the formulation and the approximations. This method shall be able to provide us an opportunity for the further study of modeling the light manipulation and the transmission enhancement with the indented double slits or related hybrid configurations.

The method could also include real metals, such as Ag, Al, Au, and Cu, for the study of the plasmonic effects. For the real metals, the plasma frequency is finite and the damping coefficient is not negligible in the visible light regime \cite{Zeman} such that the dielectric function should be taken into account. Since the method is rigorously based on the Green's theorem, it indicates that the formulation of the magnetic field will be similar to Eq. (\ref{eq12}); on the other hand, the Green's function needed for each region could be obtained from the revised method of images that considers the dielectric function \cite{Yang}. In this case, the wave propagation in the structure becomes from the PEC waveguide mode to the plasmonic waveguide mode \cite{Kim1, Gordon1} and the scattering by the openings includes both the cylindrical wave mode and the surface plasmon wave mode \cite{Lalanne, Lopez1}. In contrast to the analytical methods for the plasmonic effects that require auxiliary simulations to extract the modeling parameters of the slits and the grooves \cite{Liu, Huang}, the method could directly provide solutions to the problem without simulations.

\section{Summary}
\label{s9}
In summary, we have advanced the Neerhoff and Mur's analytical method based on Green's theorem to show that, besides a single slit, the method is capable of analyzing subwavelength structures of complex configurations, including a slit patterned with grooves, a double slit, and an indented double slit for the first time. In each configuration, the system was divided into the incident, structure, and transmission regions, where the magnetic field in each region was rigorously represented from the Green's theorem. In addition to the two types of the Green's function for the wave modes in free space and inside slit, respectively, we developed a new type for those inside a groove. With the explicit Green's functions that correspond to the structures, the analytical method yielded the solutions to the fields in excellent agreement with the FDTD simulations. In contrast to the typical mode expansion methods, we obtained the solutions entirely in real space. Since the framework is simple, extended configurations can be easily analyzed, such as a slit patterned by grooves of arbitrary dimensions and distances at the entrance and/or exit sides, multiple slits with or without patterned grooves, or slits of different widths connected in series, etc. With approximation, this method could further yield the explicit solutions to the boundary fields to demonstrate the capability of translating the wave interaction problem to the corresponding physical models. We also discussed the possibility of including the plasmonic effects in the analysis. We believe that this method will be helpful for us to gain more insight in the study of subwavelength structure properties and to extend the possibility of their applications.

\appendix
\section{Solutions to the simultaneous integral equations for the single-slit configuration}
\label{sa}
As we have successfully formulated the field representation of of the simultaneous integral equations for the light transmission through subwavelength structures of several configuration, the solution to the unknown boundary fields in the single-slit configuration in Eq. (\ref{eq31}) is introduced in this appendix. The solutions to the rest of the configurations can be yielded similarly.

Since there is no analytical solution to the simultaneous integral equations so far, a descritization method for the integrals based on Refs. \cite{Betzig, Neerhoff} is employed. In the solution, the entrance and the exit openings of the slit will be discretized with a number of uniform subintervals to represent the equations with vectors and matrices. Then, the solution can be obtained numerically by the basic linear algebra operation.

We begin with the approximation of the second-type Green's function to accelerate the convergence rate of the numerical solution. In Fig. \ref{eq02}(b), suppose that $n$ is the number of subintervals that are uniformly distributed in $[x_{s} - s, x_{s} + s]$. Since the Green's function will be part of the integrals over $x'$ in Eq. (\ref{eq31}), the high-order terms $(m \geq 1)$ can be averaged within one subinterval. We write the terms as
\addtolength{\jot}{0.8em}
\begin{align} \label{eqA-01}
T(x,x',n,{x_s},s,m) & = \cos [\frac{{m\pi (x - {x_s} + s)}}{{2s}}]\frac{1}{{\Delta x}}\int_{x' - \frac{{\Delta x}}{2}}^{x' + \frac{{\Delta x}}{2}} {\cos [\frac{{m\pi (\delta  - {x_s} + s)}}{{2s}}]d\delta } \nonumber \\
&  = \frac{{2n}}{{m\pi }}\sin (\frac{{m\pi }}{{2n}})\cos [\frac{{m\pi (x - {x_s} + s)}}{{2s}}]\cos [\frac{{m\pi (x' - {x_s} + s)}}{{2s}}],
\end{align}
where $\Delta x = 2s/n$ is the subinterval width.

Now, consider that there are $ N $ subintervals uniformly distributed at both the entrance and the exit interfaces of the slit, and the center of each subinterval is
\begin{equation} \label{eqA-02}
{x_k} =  - a + (k - \frac{1}{2})d{x} \; \; \; \: {\text{for }}k = 1, \ldots {N},
\end{equation}
where $ dx = 2a / N $ is the subinterval width. Then, Eq. (\ref{eq31}) can be represented by the vectors and matrices as the following:
\begin{subequations} \label{eqA-03}
\allowdisplaybreaks
\begin{align}
2\overset{\lower0.5em\hbox{$\smash{\scriptscriptstyle\rightharpoonup}$}} {U} \overset{i}{_b} - {\overset{\lower0.5em\hbox{$\smash{\scriptscriptstyle\rightharpoonup}$}} {U} _b} & = {{\mathbf{S}}^{\mathbf{I}}}D{\overset{\lower0.5em\hbox{$\smash{\scriptscriptstyle\rightharpoonup}$}} {U} _b}, \label{eqA-03a} \\
{\overset{\lower0.5em\hbox{$\smash{\scriptscriptstyle\rightharpoonup}$}} {U} _0} & = {{\mathbf{S}}^{{\mathbf{III}}}}D{\overset{\lower0.5em\hbox{$\smash{\scriptscriptstyle\rightharpoonup}$}} {U} _0}, \label{eqA-03b} \\
{\overset{\lower0.5em\hbox{$\smash{\scriptscriptstyle\rightharpoonup}$}} {U} _b} & =  - {{\mathbf{R}}^{{\mathbf{II}}}}D{\overset{\lower0.5em\hbox{$\smash{\scriptscriptstyle\rightharpoonup}$}} {U} _0} + {{\mathbf{D}}^{{\mathbf{II}}}}{\overset{\lower0.5em\hbox{$\smash{\scriptscriptstyle\rightharpoonup}$}} {U} _0} + {{\mathbf{S}}^{{\mathbf{II}}}}D{\overset{\lower0.5em\hbox{$\smash{\scriptscriptstyle\rightharpoonup}$}} {U} _b} + {{\mathbf{W}}^{{\mathbf{II}}}}{\overset{\lower0.5em\hbox{$\smash{\scriptscriptstyle\rightharpoonup}$}} {U} _b}, \label{eqA-03c} \\
{\overset{\lower0.5em\hbox{$\smash{\scriptscriptstyle\rightharpoonup}$}} {U} _0} & =  - {{\mathbf{S}}^{{\mathbf{II}}}}D{\overset{\lower0.5em\hbox{$\smash{\scriptscriptstyle\rightharpoonup}$}} {U} _0} + {{\mathbf{D}}^{{\mathbf{II}}}}{\overset{\lower0.5em\hbox{$\smash{\scriptscriptstyle\rightharpoonup}$}} {U} _b} + {{\mathbf{R}}^{{\mathbf{II}}}}D{\overset{\lower0.5em\hbox{$\smash{\scriptscriptstyle\rightharpoonup}$}} {U} _b} + {{\mathbf{W}}^{{\mathbf{II}}}}{\overset{\lower0.5em\hbox{$\smash{\scriptscriptstyle\rightharpoonup}$}} {U} _0}, \label{eqA-03d}
\end{align}
\end{subequations}
where the components of the vectors are given by
\begin{equation} \label{eqA-04}
{(\overset{\lower0.5em\hbox{$\smash{\scriptscriptstyle\rightharpoonup}$}} {U} \overset{i}{_b})_k} = U_b^i({x_k})
\end{equation}
with similar expression for $ {({\overset{\lower0.5em\hbox{$\smash{\scriptscriptstyle\rightharpoonup}$}} {U} _b})_{{N} \times 1}} $, $ {(D{\overset{\lower0.5em\hbox{$\smash{\scriptscriptstyle\rightharpoonup}$}} {U} _b})_{{N} \times 1}} $, $ {({\overset{\lower0.5em\hbox{$\smash{\scriptscriptstyle\rightharpoonup}$}} {U} _0})_{{N} \times 1}} $,  $ {(D{\overset{\lower0.5em\hbox{$\smash{\scriptscriptstyle\rightharpoonup}$}} {U} _0})_{{N} \times 1}} $, and the components of the matrices are
\begin{subequations} \label{eqA-05}
\allowdisplaybreaks
\addtolength{\jot}{0.8em}
\begin{align}
s_{k,j}^I & = \frac{{id{x}}}{2}H_0^{(1)}({k_0}|{x_k} - {x_j}|) \qquad \; \; \: \, {\text{for }} k \neq j, \nonumber \\
& = \frac{{id{x}}}{2}\{ H_0^{(1)}(\frac{{{k_0}d{x}}}
{2}) + \frac{\pi }{2}[{{\mathbf{H}}_0}(\frac{{{k_0}d{x}}}{2})H_1^{(1)}(\frac{{{k_0}d{x}}}
{2}) \nonumber \\
& - {{\mathbf{H}}_1}(\frac{{{k_0}d{x}}}{2})H_0^{(1)}(\frac{{{k_0}d{x}}}{2})]\} \qquad {\text{for }} k = j, \label{eqA-05a} \\
s_{k,j}^{III} & = \frac{{id{x}}}{2}H_0^{(1)}({k_0}|{x_k} - {x_j}|) \qquad \; \; \: \, {\text{for }} k \neq j, \nonumber \\
& = \frac{{id{x}}}{2}\{ H_0^{(1)}(\frac{{{k_0}d{x}}}
{2}) + \frac{\pi }{2}[{{\mathbf{H}}_0}(\frac{{{k_0}d{x}}}{2})H_1^{(1)}(\frac{{{k_0}d{x}}}{2}) \nonumber \\
& - {{\mathbf{H}}_1}(\frac{{{k_0}d{x}}}{2})H_0^{(1)}(\frac{{{k_0}d{x}}}{2})]\} \qquad {\text{for }} k = j, \label{eqA-05b} \\
r_{k,j}^{II} & = \frac{{id{x}}}{{4a{\gamma _0}}}{e^{i{\gamma _0}b}} + \frac{{id{x}}}{{2a}}\sum\limits_{m = 1}^\infty  {\frac{1}{{{\gamma _m}}}T({x_k},{x_j},{N},0,a,m){e^{i{\gamma _m}b}}}, \label{eqA-05c} \\
d_{k,j}^{II} & = \frac{{d{x}}}{{4a}}{e^{i{\gamma _0}b}} + \frac{{d{x}}}{{2a}}\sum\limits_{m = 1}^\infty  {T({x_k},{x_j},{N},0,a,m){e^{i{\gamma _m}b}}}, \label{eqA-05d} \\
s_{k,j}^{II} & = \frac{{id{x}}}{{4a{\gamma _0}}} + \frac{{id{x}}}{{2a}}\sum\limits_{m = 1}^\infty  {\frac{1}{{{\gamma _m}}}T({x_k},{x_j},{N},0,a,m)}, \label{eqA-05e} \\
w_{k,j}^{II} & = \frac{{d{x}}}{{4a}} + \frac{{d{x}}}{{2a}}\sum\limits_{m = 1}^\infty  {T({x_k},{x_j},{N},0,a,m)}. \label{eqA-05f}
\end{align}
\end{subequations}
The Struve functions $ \mathbf{H}_{0} $ and $ \mathbf{H}_{1} $ in Eqs. (\ref{eqA-05a}) and (\ref{eqA-05b}) result from an integral of the Hankel function about its singularity as outlined by Abramowitz and Stegun \cite{Abramowitz}. Note that the sizes of the matrices $ \mathbf{S^I} $, $ \mathbf{S^{III}} $, $ \mathbf{R^{II}} $, $ \mathbf{D^{II}} $, $ \mathbf{S^{II}} $,$ \mathbf{W^{II}} $ are all $ N \times N $.

In Refs. \cite{Betzig, Neerhoff}, Eq. (\ref{eqA-05f}) is implicitly simplified to be $dx/(4a)$. The simplification could be valid because, if we ignore the high-order terms ($m \geq 1$) and assume that $ U_{b}(x') $ and $ U_{0}(x') $ in Eq. (\ref{eqA-03c}) and Eq. (\ref{eqA-03d}) are invariant to the integral. Since then, the matrix operations $ {{\mathbf{W}}^{{\mathbf{II}}}}{\overset{\lower0.5em\hbox{$\smash{\scriptscriptstyle\rightharpoonup}$}} {U} _b} $ and $ {{\mathbf{W}}^{{\mathbf{II}}}}{\overset{\lower0.5em\hbox{$\smash{\scriptscriptstyle\rightharpoonup}$}} {U} _0} $ will result in  $ 1/2{\overset{\lower0.5em\hbox{$\smash{\scriptscriptstyle\rightharpoonup}$}} {U} _b} $ and $ 1/2{\overset{\lower0.5em\hbox{$\smash{\scriptscriptstyle\rightharpoonup}$}} {U} _0} $, respectively. However, it is important to keep in mind that the simplification is not able to apply for all of the configurations in this article. Hence, we keep the complete terms in the equations for the integrity of the formulation.

We can further eliminate $ {\overset{\lower0.5em\hbox{$\smash{\scriptscriptstyle\rightharpoonup}$}} {U} _0} $ and $ {\overset{\lower0.5em\hbox{$\smash{\scriptscriptstyle\rightharpoonup}$}} {U} _b} $ in Eq. (\ref{eqA-03}) to represent the original integral equations in terms of $ D{\overset{\lower0.5em\hbox{$\smash{\scriptscriptstyle\rightharpoonup}$}} {U} _0} $ and $ D{\overset{\lower0.5em\hbox{$\smash{\scriptscriptstyle\rightharpoonup}$}} {U} _b} $, and simplify them to the form:
\begin{equation} \label{eqA-06}
\left[ {\begin{array}{*{20}{c}}
   {{{\mathbf{A}}_{{\mathbf{11}}}}} & {{{\mathbf{A}}_{{\mathbf{12}}}}}  \\
   {{{\mathbf{A}}_{{\mathbf{21}}}}} & {{{\mathbf{A}}_{{\mathbf{22}}}}}  \\

 \end{array} } \right]\left\{ {\begin{array}{*{20}{c}}
   {{{\overset{\lower0.5em\hbox{$\smash{\scriptscriptstyle\rightharpoonup}$}} {B} }_1}}  \\
   {{{\overset{\lower0.5em\hbox{$\smash{\scriptscriptstyle\rightharpoonup}$}} {B} }_2}}  \\

 \end{array} } \right\} = \left\{ {\begin{array}{*{20}{c}}
   {{{\overset{\lower0.5em\hbox{$\smash{\scriptscriptstyle\rightharpoonup}$}} {C} }_1}}  \\
   {{{\overset{\lower0.5em\hbox{$\smash{\scriptscriptstyle\rightharpoonup}$}} {C} }_2}}  \\

 \end{array} } \right\},
\end{equation}
where the matrices and vectors are
\begin{subequations} \label{eqA-07}
\allowdisplaybreaks
\begin{align}
{({{\mathbf{A}}_{{\mathbf{11}}}})_{{N} \times {N}}} & = {{\mathbf{D}}^{{\mathbf{II}}}}{{\mathbf{S}}^{{\mathbf{III}}}} - {{\mathbf{R}}^{{\mathbf{II}}}}, \label{eqA-07a} \\
{({{\mathbf{A}}_{{\mathbf{12}}}})_{{N} \times {N}}} & = {{\mathbf{S}}^{{\mathbf{II}}}} - ({{\mathbf{W}}^{{\mathbf{II}}}} - {\mathbf{I}}){{\mathbf{S}}^{\mathbf{I}}}, \label{eqA-07b} \\
{({{\mathbf{A}}_{{\mathbf{21}}}})_{{N} \times {N}}} & = ({{\mathbf{W}}^{{\mathbf{II}}}} - {\mathbf{I}}){{\mathbf{S}}^{{\mathbf{III}}}} - {{\mathbf{S}}^{{\mathbf{II}}}}, \label{eqA-07c} \\
{({{\mathbf{A}}_{{\mathbf{22}}}})_{{N} \times {N}}} & = {{\mathbf{R}}^{{\mathbf{II}}}} - {{\mathbf{D}}^{{\mathbf{II}}}}{{\mathbf{S}}^{\mathbf{I}}}, \label{eqA-07d} \\
{({\overset{\lower0.5em\hbox{$\smash{\scriptscriptstyle\rightharpoonup}$}} {B} _1})_{{N} \times 1}} & = D{\overset{\lower0.5em\hbox{$\smash{\scriptscriptstyle\rightharpoonup}$}} {U} _0}, \label{eqA-07e} \\
{({\overset{\lower0.5em\hbox{$\smash{\scriptscriptstyle\rightharpoonup}$}} {B} _2})_{{N} \times 1}} & = D{\overset{\lower0.5em\hbox{$\smash{\scriptscriptstyle\rightharpoonup}$}} {U} _b}, \label{eqA-07f} \\
{({\overset{\lower0.5em\hbox{$\smash{\scriptscriptstyle\rightharpoonup}$}} {C} _1})_{{N} \times 1}} & = 2({\mathbf{I}} - {{\mathbf{W}}^{{\mathbf{II}}}})\overset{\lower0.5em\hbox{$\smash{\scriptscriptstyle\rightharpoonup}$}} {U} \overset{i}{_b}, \label{eqA-07g} \\
{({\overset{\lower0.5em\hbox{$\smash{\scriptscriptstyle\rightharpoonup}$}} {C} _2})_{{N} \times 1}} & =  - 2{{\mathbf{D}}^{{\mathbf{II}}}}\overset{\lower0.5em\hbox{$\smash{\scriptscriptstyle\rightharpoonup}$}} {U} \overset{i}{_b}. \label{eqA-07h}
\end{align}
\end{subequations}
Since $\mathbf{A}$ and $\overset{\lower0.5em\hbox{$\smash{\scriptscriptstyle\rightharpoonup}$}} {C}$ are composed completely of the known components, the unknowns $D{\overset{\lower0.5em\hbox{$\smash{\scriptscriptstyle\rightharpoonup}$}} {U} _0}$ and $D{\overset{\lower0.5em\hbox{$\smash{\scriptscriptstyle\rightharpoonup}$}} {U} _b}$ can be readily determined. Substitution them into Eq. (\ref{eqA-03}) then determines the remaining unknowns ${\overset{\lower0.5em\hbox{$\smash{\scriptscriptstyle\rightharpoonup}$}} {U} _0}$ and ${\overset{\lower0.5em\hbox{$\smash{\scriptscriptstyle\rightharpoonup}$}} {U} _b}$. To summary, by formulating the problem in the language of linear algebra, a discrete solution can be found for the complete wave interaction problem. As the number of subintervals $N$ increases, this discrete solution should approach the true continuous solution.

\bibliographystyle{unsrt}  
\bibliography{arXiv-Method}  

%
%
%
%

\end{document}